# Event-driven agility of interoperability during the Run-time of collaborative processes

Anne-Marie Barthe-Delanoë[a*], Sébastien Truptil[a], Frédérick Bénaben[a], Hervé Pingaud[b]

[a]*Mines Albi-Carmaux, Campus Jarlard, 81013 Albi CT Cedex 9, France*
{anne-marie.barthe,sebastien.truptil,frederick.benaben}@mines-albi.fr

[b]*Université Jean-François Champollion, Place de Verdun, 81012 Albi, France*
herve.pingaud@mines-albi.fr

*Corresponding author (+33 5 63 49 31 38)

ABSTRACT

The modern business environment tends to involve a large network of heterogeneous people, devices and organizations that engage in collaborative processes among themselves. Given the nature of this type of collaboration and the high degree of interoperability between partner Information Systems, these processes need to be agile in order to respond to changes in context, which may occur at any time during the collaborative situation.

The objective is to build a Mediation Information System (MIS), in support of collaborative situations, whose architecture must be (i) built to be relevant to the collaborative situation under consideration, (ii) more easily integrated into the existing systems, and (iii) sufficiently agile, through its awareness of the environment and of process events, and through the way it reacts to events detected as being relevant.

To apply agility mechanisms, it is crucial to detect the significant events that will lead to a subsequent evolution of the situation (detection step). Event-Driven Architecture (EDA) is used to design the structure of the part of the system that is in charge of MIS agility. This architecture takes the events into account, manages them and, if needed, uses them to trigger the adaptation of the MIS.

We have defined a means to monitor the evolution of the situation. If relevant changes are detected, and if the situation does not evolve in the expected way, an adaptation is proposed.

It is concluded that the principles of detection and adaptation, combined with the responsiveness of the system (provided by the automation of transitions), and based on Event Driven Architecture principles, together provide the agility required for collaborative processes.

**KEYWORDS**

*Complex event processing   Agility  Adaptation    Decision making          Dynamic change          Interoperability*

## 1. Introduction

Nowadays, organizations (such as enterprises, institutions or administrations), the people who work in them and the devices they use, all have to work together and take part in collaboration to be able to operate in an unstable environment. This need for interconnection, and more precisely for collaboration, is revealed by contexts as numerous and various as social networking, domotics, business partnerships, subcontracting, or crisis situations. Our environment is thus tending to become a large network of people, machines and organizations (i.e. the collaborative partners), all involved in collaborative processes among themselves. But taking part in a collaborative process is not necessarily easy for the partners, especially in a context of ephemeral collaboration. Moreover, industrial relationships have evolved and they are no longer based on long-term collaboration. Today they are also based on opportunistic collaboration, rapidly established and dissolved. In this context, the notion of agility has emerged with the understanding that collaboration needs to be flexible.

The ability to collaborate with clients, providers or even competitors has always been a critical requirement in our modern multi-organizations-based ecosystem [13]. However, if collaborating used to concern closely-related organizations (from a geographical point of view), and required time to define a stable and durable relationship, this is no longer the case: nowadays, organizations need to establish their - potentially short-lived - collaborations with partners from all around the world, in a very reactive way in order to seize very fleeting business opportunities. It can be argued that the business ecosystem has evolved from a strongly crystallized structure into a very fluid environment. In this free-flowing context, collaborating is more a way to seize opportunities and to stay dynamically on the top of the wave, rather than a structuring element defining the intensity of the organizations' integration in their geographical and business environment.

Furthermore, Information Systems (ISs) can be considered, on one hand, as the functional backbone of organizations [41] (insofar as they assume the management of their information, functions and behavior) and on the other hand, as the main interface (the visible part of the organization as described by Morley [33]) with any potential partner. Consequently, the management of organizational collaboration should definitely aim to achieve information system interoperability. Our starting point is to approach the collaboration issue through IS interoperability, thus satisfying the business requirements of the organizations.

In this article, we propose an approach and a set of theoretical results to support collaboration (i.e. the collaborative processes) and enhance its agility. Regarding the specific research works presented in this article, the overall contribution is the following: [5] did define the precise context and requirements of this agility feature while the current article is in charge of providing the reader with all the theoretical studies and results to meet these requirements. Consequently, the contribution of this article mainly concerns the theoretical definition of an agile framework for a Mediation Information System (MIS) (that has been described in previous works).

The remainder of this article is organized as follows: Section 2 gives an overview of the literature on related products and research projects. This section also presents some considerations regarding agility. Section 3 presents and describes our proposal of a platform to support collaboration and to ensure the agility of the processes. Section 4 contains a discussion about the findings, suggestions for further work, and a conclusion.

## 2. Background

We first provide a brief background to collaboration support tools and flexibility (agility) principles, presenting several commercial and research works on workflow agility. Then, some core ideas of Event-Driven Architecture (EDA) are presented, to justify the need for such architecture and the use of a Complex Event Processing (CEP) engine in the platform.

### 2.1. Tools to support collaboration and its agility

For a decade, several commercial products and research projects have been attempting to design, orchestrate and provide agility to collaborative workflows. On the commercial side, the major actors are Bonita and the tools that are based on Architecture of Integrated Information Systems (ARIS) [47]. Bonita Open Solution (developed by Bonitasoft [8]) offers a suite of tools to design, execute and monitor processes. ARIS tools aim to model enterprises.

Generally, there are platforms providing functions to model the business processes and to implement them as workflows, to execute and monitor them. The ARIS approach can also integrate the notion of events inside the process modeling. An interesting point here is ARIS' ability to combine determined process fragments according to received events. In a way, the ARIS approach manages workflow adaptation (but in a determinist manner).

We can cite the WORKPAD project [11], which designed and developed a software infrastructure to support collaboration in emergency/disaster scenarios. This project aimed to create communities of Public Safety Systems (PSSs) and to enable mobile teams to exploit PSSs through mobile technologies, process management and geo-collaboration. On the adaptation side, they focused on recovering the disconnecting nodes through specific tasks. The CRISIS [54] project aimed at developing a train-on-demand simulation platform to train first responders and crisis managers: their platform helps to explore decision-making under conditions of uncertainty. They do not really orchestrate workflows: they focus more on the decision-making part when facing new risks, or new uncertainties.

Other platforms propose event subscription and publication. For example, we can cite the Pachube project [21], which offers a platform to subscribe to and publish events. But Pachube does not offer any computation on them. The PRONTO project [30] aims at collecting and deducing complex events from event streams, but it does not focus on the workflow management part.

The European project PLAY proposes a modeling framework named SANs (Situation Action Networks [51]. SANs are goal-directed tree models that allow to find alternative activities to reach the goals defined by the collaborative processes. The moment of the choice to adapt or not the processes is based on determined milestones.

The following table presents these existing results regarding *agility* of collaborative workflows and mainly according to three main components of agility (to be defined more precisely in next section 2.2): *detection* of a need of adaptation, *adaptation* of workflows and *responsiveness* of the whole. The first feature (*detection*) concerns the ability of the product/project to diagnose that the currently running behavior is no longer in line with the situation (for any known or unknown reason). The second feature (*adaptation*) concerns the ability of the product/project to define (on the fly) a new and relevant behavior (*i.e.* collaborative workflows) according to the knowledge provided by the *detection* feature. Finally the third feature (*responsiveness*) concerns the ability of the product/project to perform *detection* and *adaptation* in a fast and reactive way (in order not to get a "slow motion reconfiguration", which would definitely not ensure real-time agility).

**Table 1** Overview of existing solutions to provide agility to collaborative workflows.

| Product/Project | Detection of a need for adaptation | Adaptation of workflows | Responsiveness |
| --- | --- | --- | --- |
| Bonita | No | No | No |
| ARIS | Yes (automated, event-driven) | Yes (automated and pre-determined alternatives) | N/A |
| TIBCO | Yes (manually done) | Yes (manually done) | No |
| WORKPAD | Yes | No | Yes |
| CRISIS | Yes | Yes (partial adaptation) | Yes |
| PRONTO | N/A | No | Yes |
| PACHUBE | Yes | No | No |
| PLAY | Yes (pre-determined milestones) | Yes (pre-determined alternatives) | Yes |

Table 1 shows us that, for the moment, there are no commercial products or research projects that propose a platform encompassing all the functions of collaborative process design, which can run them, make them context-aware and then adapt them in a short time.

*2.2. Concepts of agility*

The notion of agility has been widely discussed. As an introduction, the Collins dictionary defines agility as the power of moving quickly and easily. For Badot [4], agility is a reconfiguration of the system to satisfy a need for adaptation. For other authors, such as Kidd [23], Lindberg [25] and Sharifi [49], agility is a need for flexibility, responsiveness or adaptability. In logistics, flexibility is seen as "the ability to meet short-term changes" [50] and is differentiated from adaptation over time in response to a change [31].

Considering the notions of responsiveness (related to the speed of adaptation), adaptation (related to the magnitude of this adaptation) and detection (related to the moment of adaptation), we propose the following definition of agility: *agility is the ability of a subject to lead as quickly as possible, on the one hand, to the detection of its mismatch to a given context, on the other hand, to the setting up of the required adaptation*. In our context, this means that we need to detect when a workflow is not relevant with regard to the collaborative goals and the current context of the collaborative situation (detection), and what needs to be done to deal with this issue (adaptation), as fast as possible (responsiveness).

Workflow adaptation approaches are various: we can mention in particular the adaptive inter-organizational workflows of Andonoff [2], the ADEPTflex approach of Reichert [43] and, along the same lines, the research works of Van der Aalst [1]. Recently, Schonenberg [48] has proposed a taxonomy of workflow flexibility approaches (presented in Figure 1) that shows four main classes of approach:

- Flexibility by design: this provides flexibility in the process design by including numerous alternative execution paths that can cover as many as possible of the different opportunities, instabilities and threats associated with the behavioral dynamics of the studied situation. The selection of the most appropriate branch is performed at Run-time. For instance, this is the approach proposed by Rüppel [45] in his research work to define many possible response processes to a crisis situation, such as a flood in Germany.

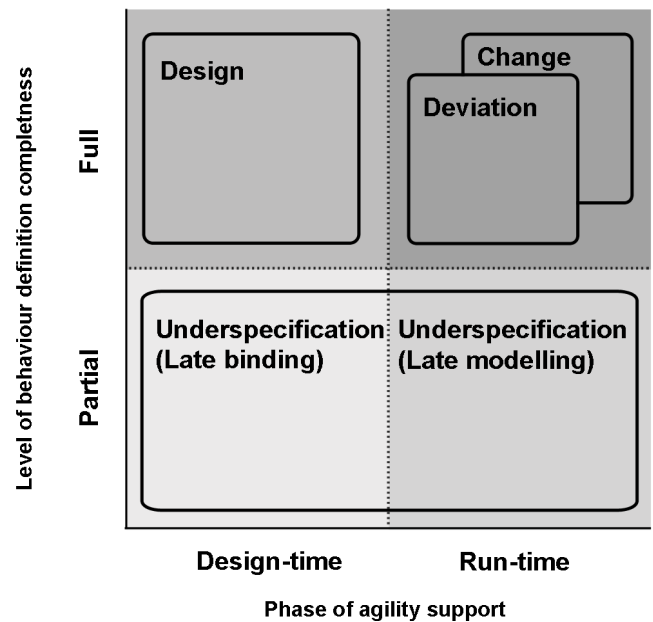

**Fig. 1.** The four approaches of agility, according to Schonenberg et al. [48].

- Flexibility by deviation: this kind of flexibility is provided during Run-time. It allows the order of the execution of activities to be changed, without changing the activities themselves. In other words, it allows a task to be cancelled, restarted or skipped. For example, the Flow system described by Van der Aalst [48] can use flexibility by deviation.
- Flexibility by underspecification: this approach partially defines the processes during the Design-time, and completes them during the Run-time. This kind of flexibility is based on the fact that tasks or abstract sub-processes cannot be defined with precision, but they can be identified at Run-time (i.e. once some choices have been made). This kind of flexibility is supported by the YAWL system. This approach is sub-divided as follows:
  - Late binding: in this concept, the elements of the workflows are viewed as objects whose implementation is defined during the process Run-times [14] [18] [20]. The user is left to choose among a set of Run-time options at the appropriate moment,
  - Late modeling: here, some elements of the workflows are not identified during the Design-time, but are specified during the Run-time of the processes [44]. This option has to allow the execution (during the Run-time) of the Design-time tools in order to add parts to the incomplete processes,
- Flexibility by change: this approach aims at changing the definition of running processes, by inserting or deleting tasks. This approach is the most commonly used, for example in ADEPTflex [1][9], or in the research works of Casati [10], Sadiq [46] and Weske [52].

In our research work, we position our adaptation proposal by mixing flexibility by deviation and flexibility by change, to define an ad-hoc approach, where the behavioral model of the collaborative situation is defined at the point of need and on-the-fly.

*2.3. Implementation of agility*

As stated in Section 1, the collaborative environment is not static; it is constantly changing. To maintain their collaboration relevant at time t, the partners have to identify and react to certain situations as they occur. These situations can be either positive or negative with respect to the collaborative goals.

Thus any changes, any evolution, any information that could challenge the accuracy and relevance of the collaborative process need to be managed, as Rao [42] underlines it.

According to Chandy and Schulte [12], Etzion and Niblett [17], and Luckham and Schulte [28] this or that occurrence and any particular embedded data can be considered (and managed) as events. They are produced by:

- The people in the collaborative situation and their machines,
- The services used by the collaborative workflows.

An event-driven approach will allow our MIS to monitor the changes as they happen, in a real-time perspective.

In the literature, the combination of both Event-Driven Architecture (EDA) principles and SOA principles has been widely discussed. We agree with the view of authors like Josuttis [22], Luckham [26][27], Maréchaux [29] and Michelson [32] who affirm that EDA should not be seen as a competitor to SOA but as additional principles to complement SOA principles.

As our MIS is SOA-based, the addition of an EDA layer allows us to gather knowledge about the collaborative situation, the collaborative environment and the collaborative process in almost real-time by taking events into account, thus making our MIS context aware.

But EDA is not only about managing event exchanges between processes, people and machines: it also concerns the business level by filtering and applying business rules to detect relevant events or combinations of events. For example, two events (called simple events), which are not seen as risks or opportunities when viewed separately, may have a different meaning if they are considered together (and so they create a complex event). This is called Complex Event Processing (CEP), which is technically carried out by a CEP engine [30].

Moreover, the analysis of a large amount of different events (including the interactions between these events) may become a very complex task if it is done only by human beings, with regard to the time criteria and the available human resources. So the MIS also needs to take charge of the analysis of these different events in addition to simple event detection. In fact, the MIS not only detects simple events but also events resulting from a combination of simple events (these complex events are detected through the execution of defined business rules among simple events). All these events will be used by the agility service of the MIS, in order to detect a possible mismatch between the situation in the field of the collaboration and the executed collaborative processes at time t.

Another interesting point concerning EDA is its ability to provide very loose coupling between applications (i.e. web services in our case) through the publish/subscribe mechanism (as described in [35]: applications subscribe to a certain type of event (or pattern of events) and not to a specific source of event. We can imagine that two web services provide the same kind of information, through the publication of the same event type *A*. Other web services subscribe to the event type *A*. If a third provider enters the network and publishes events of event type *A*, the subscribers will receive them without any technical modification or interface creation. This ability fits completely with our need for an agile structure.

### 3. Our proposal

#### 3.1. Overview of proposal

According to the European Network of Excellence, InterOp, interoperability is "*the ability of a system or a product to work with other systems or products without special effort from the customer or user*" [24]. It is also defined by Pingaud [39] as "*the ability of systems, natively independent, to interact in order to build harmonious and intentional collaborative behaviors without deeply modifying their individual structure or behavior*".

As it seems that the ISs of the partners involved cannot natively assume the functions of data transfer and translation, and of service management and collaborative workflow orchestration (except with high technical standardization constraints, such as creating specific interfaces between partner applications or having to recode applications, none of which fits the given definition of interoperability), we have to find a way to support these collaborative functions. Figure 2 presents the three architectural alternatives.

Case (a) is based on a peer-to-peer approach: it is a highly-coupled architecture, with technical constraints which can be very expensive in terms of the dynamic context of the collaborative situation and the time required to set up such collaborative architecture.

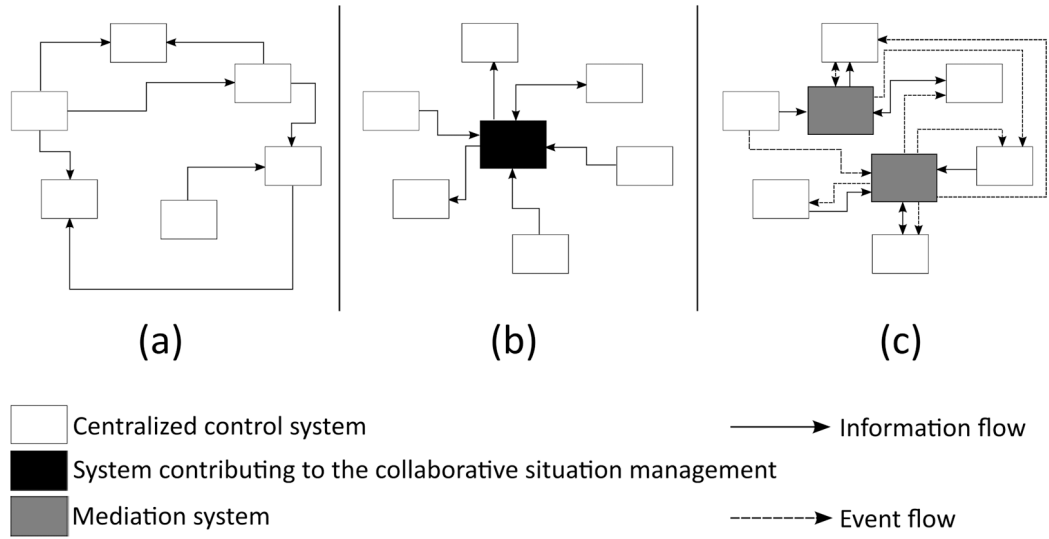

**Fig. 2.** Network architecture alternatives: from centralization to mediation.

On the contrary, case (b) proposes a mediation approach (as defined by Wiederhold [53]): this is a low-coupled architecture (taking account of the constraint of evolution in the collaborative process), with an IS orchestrating all the exchanges. A Mediation Information System (MIS), based on Service-Oriented Architecture (SOA) principles –a third-party system in charge of the coordination of the partners' activities according to collaborative processes– is a credible and pertinent way to support IS interoperability (as detailed in [7]). The Mediation Information System Engineering (MISE) 1.0 project (2004-2008) has been successfully completed. Its aim was to design and develop such a MIS [5]. The MISE 2.0 project aims at solving some assumptions made during the MISE 1.0 project.

Finally, case (c) proposes a distributed mediation approach, using communication through event flows. This kind of architecture is based on both SOA principles and Event-Driven Architecture (EDA) principles (detailed in the Section 2.3 of this paper) and complements the previous approach by the addition of an EDA layer.

The approach of the MISE project provides several improvements in the way collaboration can be managed. It relies on several concepts, paradigms and theories combined together to provide various benefits. Schematically, these benefits may be presented according to the following list:

- MISE aims at ensuring heterogeneous IS interoperability thanks to a mediation information system (MIS): this first component of the approach provides the network with a theoretical structure that ensures three functions of interoperability: information translation, functions sharing and collaborative processes orchestration.
- MISE is based on a Model-Driven Engineering approach (MDE): this second characteristic ensures a high level of automation (thanks to automated model transformations) and the use of gathered, imported or generated knowledge at the accurate abstraction level.

- MISE uses also a Business Process Modeling approach (BPM): this third element ensures that the generated and implemented behavior covers the whole business domain under consideration through relevant structured process cartography, and meets the collaborative requirements.
- MISE uses an Enterprise Service Bus (ESB) to deploy a Service-Oriented Architecture (SOA): this fourth characteristic supports a high level of connectivity and, combined with a workflow engine, brings orchestration ability. Furthermore, such an architecture provides a platform to merge Design-time and Run-time by connecting Design-time services and Run-time services on the same ESB thus integrating the adaptability feature.
- The fifth feature of MISE is the exploitation of an Event-Driven Architecture (EDA) in order to supervise collaborative behavior: this feature is based on the proposal for the agility of the collaborative processes presented in this article. It ensures that the Run-time knowledge may be instantaneously collected in order to be exploited: first for choreography by transmitting information and secondly, for the detection of evolutions that require adaptation by diagnosing that the field situation (represented by events sent by the field of the collaboration) is no longer in line with the expected situation.

The several layers of the MIS defined by the MISE project —collaborative situation characterization, collaborative process cartography and workflow implementation— will be briefly explained at the beginning of Section 3.2, in order to facilitate the understanding of our proposal concerning workflow agility.

### 3.2. Details of our proposal

The MISE 2.0 project design approach is based on a Model-Driven Approach (MDA) through an automated model transformation including (as shown on Figure 3) two major parts: Design-time and Run-time. The Design-time may be compared to a mountain canyon waterfall with three potholes following each other.

The water is jumping in a one-way direction from one pothole to another one —representing the steps of characterization, cartography deduction and implementation— and embeds new characteristics such as temperature, color, sediments, etc. which represent the data and the knowledge obtained at the end each steps of the Design-time through model transformation or extraction of technical settings.

The end of the waterfall represents the deployment and the execution of the workflow based on the last step of the Design-time, while a one-way trail on the side of the waterfall (Agility service) allows to go back to any desired pothole (top, middle or bottom pothole), depending on the analysis of the data gathered at the end of the waterfall. These two steps constitute the Run-time.

#### 3.2.1. Design-time

First of all, the knowledge about the collaborative situation is gathered (situation layer). Mu [34] proposes a metamodel of the collaborative situation. A modeling tool (Mediator Modeling 2ool), based on the metamodel, has been designed to support objective and function (service) modeling. A model of the collaborative situation is built.

Secondly, the collaborative process cartography is automatically deduced through the Mediator Modeling 2ool which supports a Collaborative Business Process Deduction methodology [34].

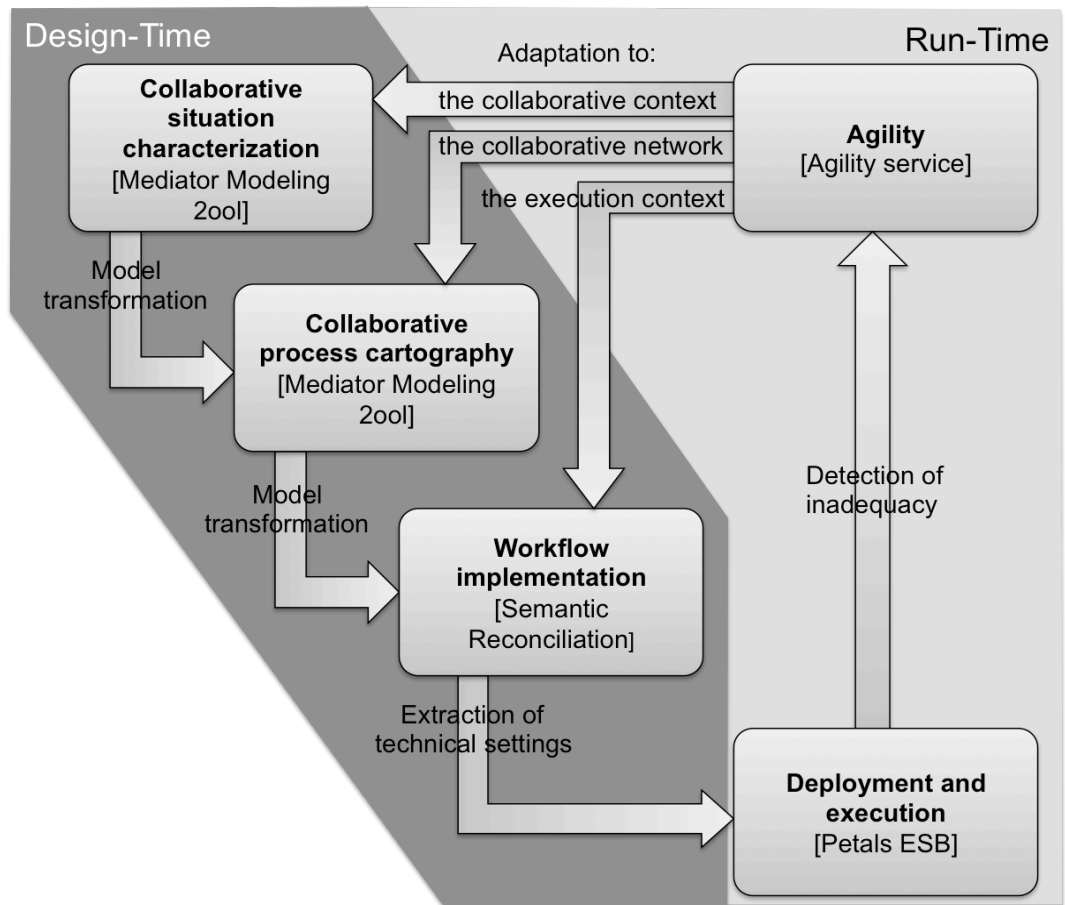

**Fig. 3.** Overview of the MIS architecture and tools.

This methodology aims at (i) selecting the business services corresponding to the collaborative objectives; (ii) ordering the business services to obtain a BPMN collaborative process cartography. Step (i) is supported by a collaborative ontology fulfilled with instances from specific domains and an Objective-Function Mapping algorithm. Step (ii) is based on model transformation rules and mediation concepts ontology.

The last step of our MIS's Design-time is its deployment. There is a focus on the analysis of semantic issues during the transformation of a BPMN collaborative process into a runnable workflow for a specific target platform that is an Enterprise Service Bus (ESB) [6]. Three semantic gaps to obtain a concrete workflow from the abstract level are identified:

- Matching information and data (informational semantic issue).
- Matching business activities and technical services (functional semantic issue).
- Matching business process and workflow (behavioral semantic issue).

For the moment, a specific solution is proposed for the semantic reconciliation (as it is dedicated to crisis management) where additional knowledge helps to avoid semantic problems, as there is a direct matching between business activities and technical services. The investigation for more generic solutions – where semantic knowledge is added at the abstract and technical levels (supported by existing Semantic Web Services standards, as explained in [6]) – is in progress.

### 3.2.2. Run-time

We have chosen Petals ESB, developed by the French Open Source software editor Petals/Linagora, to deploy our MIS. Petals ESB embeds a Business Process Execution Language (BPEL) workflow engine that executes the deduced collaborative workflows (Run-time), but also hosts (as web services) the two tools used to design our MIS (Design-time) (cf. Section 3.2.1): Mediator Modeling 2ool and Semantic Reconciliation.

## 3.3. Agility

This proposal clearly illustrates our will to define a bottom-up approach dedicated to the design of a mediation information system at the level of the Design-time.

Nevertheless, the notion of collaboration also presents strong constraints at the Run-time level. Indeed, the operational dynamics of collaboration may be exposed to some unanticipated unknowns that can require an evolution of the MIS. As explained in [40], there are two kinds of sources of adaptation:

- The evolution of the collaborative situation itself: the perceived characteristics of the collaboration, in particular the issues to be solved, are not the same at the beginning of the collaboration and need new collaborative workflows.
- The evolution of the collaborative workflows: the management of the collaborative workflows may evolve due to (i) a change in the structure of the collaborative partners (e.g. arrival or leaving of partners, lack of resources), (ii) a dysfunction of the execution of a service (leading to the interruption of a workflow), or (iii) an incomplete initial definition of the collaborative processes.

Moreover, this dynamic is an intrinsic component of the concept of organization collaboration in the current environment (cf. Section 1). The MIS's agility during the Run-time is a requirement that generates a certain number of repercussions on the MIS architecture, as we consider agility to be the combination of detection and adaptation, surrounded by responsiveness. It leads to define an Agility Service, in charge of the agility of the MIS.

### 3.3.1. Detection solutions

In this subsection, we will focus on how to detect when a workflow is not relevant to the collaborative goals and the current context of the collaborative situation.

Given that we have a model of the collaborative situation (cf. Section 3.2.1) and that the collaborative workflows are deduced on the basis of this model at time 0, we can say that if the model of the collaborative situation at time t has evolved, the deduced collaborative workflows may not be relevant to this situation at time t. We can also say that if the collaborative workflows do not meet the collaboration goals (or more precisely, the expected results of activities), this may be due to a change in the collaborative partners, in terms of the two kinds of sources of adaptation.

So, the detection step consists in identifying the evolutions of the collaborative situation model regarding the context at time t.

As explained in Section 2.3, we have added an EDA layer to the MIS architecture. The partners' web services involved in collaborative workflows are able to send and receive events, and (in our case) they implement the WS-Notification standard to manage the publish/subscribe mechanism [35][36][37]. For example, a web service can send events about its state (invoked, in progress, completed, interrupted) in addition to the regular message exchanges with the other web services. There are also the events sent by sensors or other devices in the field of the collaborative situation. In order to collect all the events concerning the collaborative situation and to be able to deduce new events, the CEP engine subscribes to the event types concerning the events coming from both collaborative workflow execution and the field of the collaboration. Then, the CEP emits all the gathered and created events.

The main idea here is that events give us information about the evolution of the collaborative situation in two ways:

- Some inform us about the state of the activities of the workflows (and thus their execution). For instance, in a crisis situation, the activity "extinguish fire by dumping two tanks of water by water-bombers" sends an event when it is done, so when we receive this event, we can consider that the fire is extinguished (and thus, a part of the crisis situation is solved),
- Others inform us about the reality in the field of the collaboration. For instance (again in a crisis situation), the activity "extinguish fire by dumping two tanks of water by water-bombers" is well accomplished, but the firemen in the field (in charge of this fire) report to the MIS that the fire is not completely extinguished and the wind speed is increasing. So the response (i.e. the executed workflow) to this part of the crisis situation partially failed.

We can use these events to track the changes inside the collaborative situation model by this method (illustrated by Figure 4):

- First, we duplicate the initial model of the collaborative situation (i.e. model at time 0),
- Then, we update both models with the received events (the MIS Agility Service subscribes to all the events emitted by the CEP). We obtain two models through this update:
  - The *expected model*: the planned and expected situation model at time t (i.e. what we expect to obtain when we apply the collaborative workflows to the collaborative situation). It is obtained by updating the initial model with monitoring events.
  - The *field model*: the real situation model of the collaboration at time t, whatever the applied collaborative workflows are (i.e. the "what actually happened" situation at time t). It is obtained by updating the initial model with events coming from the field.
- At time t (arbitrary chosen), we measure the divergence $\partial$ between the expected model and the field model.

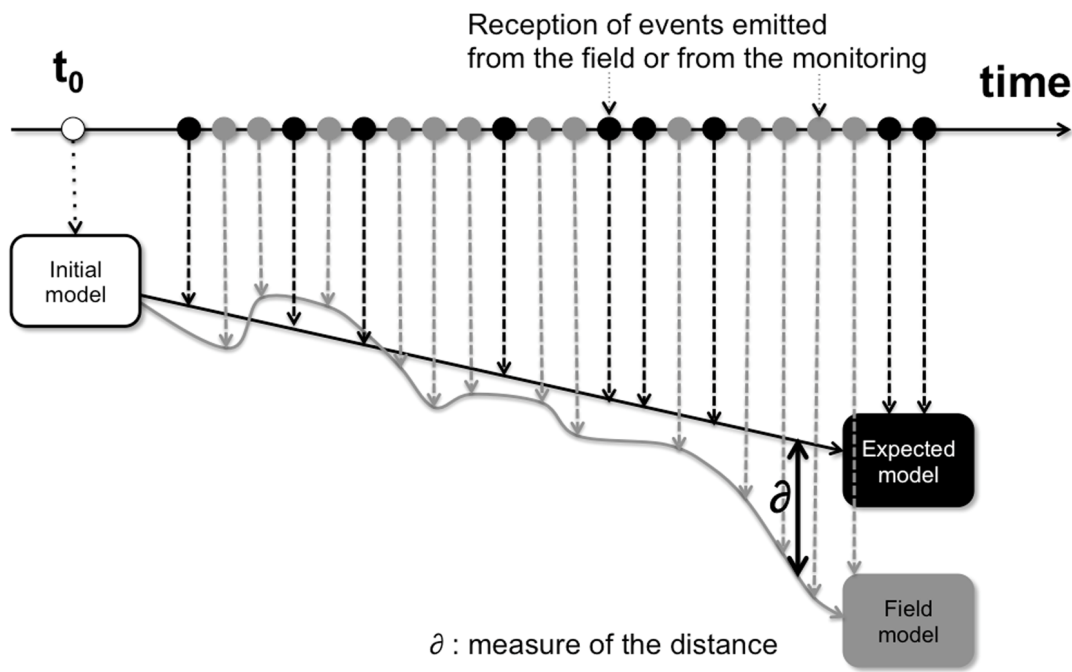

**Fig. 4**. Principles of detection of the divergence.

The measure of $\partial$ is automatically made on the whole set of points of our models in order to determine the nature of the divergence, its size and its origin. These points are the instances of the concepts described in the collaboration model. They are necessary to help to decide if an adaptation is necessary (according to business rules, defined by the collaborative partners), and if so, to give a certain number of recommendations on how to adapt the workflows.

We have explored several ways to calculate the divergence. For example, as our models are XML based, a possible approach could have been the use of algorithms for XML tree comparison (like those presented in [15] and [38]) to detect the changes, their origin and their nature. But these algorithms do not really meet our requirements which are looking for similarity (the order of nodes which are siblings does not matter in our comparision) and for a full report of the detected differences. Finally, we adapt a tool used to check the quality of XML transformations called XMLUnit [3][19] as it fits almost all our needs.

We have also taken into account the cost of the operations to be carried out (add, delete, modify) on one tree to make it similar to the other as it is a criteria to determine the size and the nature of the divergence.

The measure of $\partial$ is given by the following Formula 1:

$$\partial = \Sigma w_i . \partial_i \qquad (1)$$

$\partial_i$ is a value (between 0 and 1) representing the "cost" of instance number i of the model. If any difference is identified between expected model and field model regarding this particular instance, then this "cost" is added to the global sum (balanced with wi). By default, each instance got the "cost" 1.
$w_i$ is the predetermined weight depending on (i) the type of concept concerned by the identified difference (e.g. partner, risk, resource, activity, etc.) and (ii) the kind of difference – called "operation" here (added, deleted, updated). This weight is used to qualify each detected difference, as each difference, even on the same instance, has not the same impact on the relevancy of the processes. For example, the addition of a risk (respectively, the deletion of a partner) has more negative impact on the processes than the deletion of a risk (respectively, the addition of a new partner). These weights wi are predetermined depending element types and operations.
$\partial$ is automatically calculated by the Agility Service of the MIS defined through the MISE project. Partners also define a threshold. If $\partial$ is over this threshold, the Agility Service will automatically move to the adaptation step.

As we know how to detect the possible divergence between the expected situation (regarding the applied collaborative workflows) and the situation in the field, we can focus on the ways to accomplish the associated adaptation.

### 3.3.2. Adaptation solutions

We propose combining two kinds of system adaptations (to the most appropriate conformation): (i) the ability to evolve in a predetermined closed geometry and/or (ii) the ability to redesign a new structure fitting the situation.

The first point refers more to a "Design-time" agility, based, for instance, on risk studies and leading to the building of models including a number of conditional branches to optimize coverage of the possibilities. The second point refers more to a "Run-time" agility where the (re)building of the best possible conformation has to be improvised, at the convenient moment.

In the context of the problem discussed here, it is mainly a question of allowing the interruption of the orchestration of the collaborative process(es) at any time, in order to call – during the Run-time – the tools of the Design-time to redesign the process(es) in a more appropriate way. Thus this last option would allow a return to the most relevant level of abstraction regarding the desired adaptation (characterization of the collaborative situation, modeling of the dynamic of the response, definition of the mediator, deployment of the MIS) and thus to nest, or even to merge, the Design-time and the Run-time. We have chosen this option: the MIS is SOA based. The Design-time tools are hosted on the ESB as web services provided by the MIS, which facilitates the return to the design step (to carry out the adaption) during the Run-time, as shown in Figure 4. The MIS can go back to Design-time at the three levels (described in Section 3.2):

- Characterization of the collaborative situation (point 1 on Figure 5),
- Process cartography design (point 2),
- Mediation Information System deployment (point 3).

The MIS then executes the adapted workflows that call the web services provided by the partners and the devices (point 4).

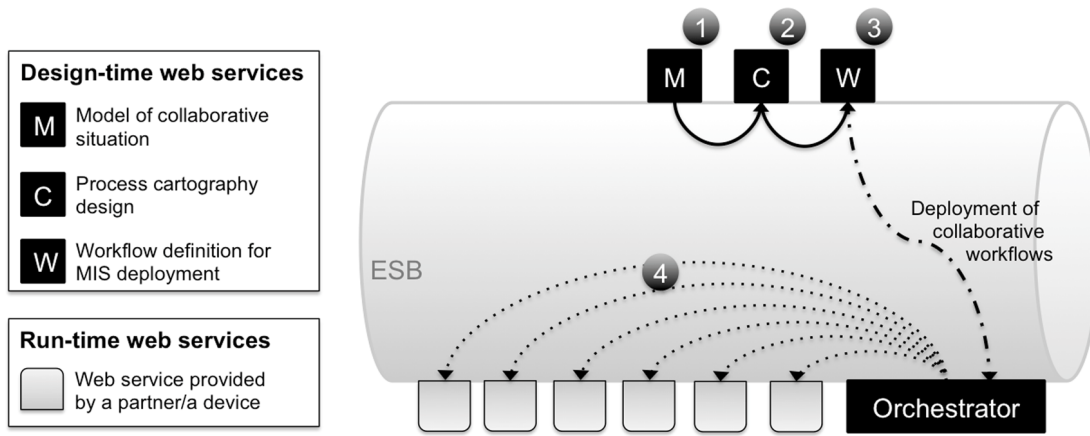

**Fig. 5.** An overview of the MIS architecture in the MISE 2.0 project.

## 4. Perspectives and conclusion

The implementation of the EDA layer and the set of agility tools is currently in progress. As regards the EDA layer, this part has been completed: the web services are able to send events (producer role) and to subscribe to events (consumer role) through the implementation of the WS-Notification standard. A CEP engine (Esper, developed by the American software editor EsperTech [16]) is also integrated in the architecture: it subscribes to various event types and applies determined business rules on events in order to create new complex events (as described in Section 2.3). It is also able to emit events following the WS-Notification standard: a specific software component has been developed to allow web services to subscribe to the CEP events.

Two major stakes are noticeable according to these research works:

- The first one concerns the coordination of partners: it is easy to imagine that partners of a collaborative situation are ready to perform efficiently their tasks by mobilizing their own capabilities, however, it is also important to notice that cultural, technical or practical heterogeneity of partners may engender friction. Deploying a MIS could compensate this issue. This is the first benefit of these research works.
- The second one concerns the management of data, and especially "big data". Any field (industrial field, battle field, public field, crisis field, etc.) is going to be more and more data provider. People, devices, building, networks, and so on, are about to generate more and more numerical data (through sensors, social networks, etc.). This is obviously an improvement according to the "management" point of view, but it definitely requires systems able to deal with this amount of data. Combining, analyzing, comparing and exploiting this data is no longer a human-sized task; the agility management mechanism described in this article is a way to ensure that layer between "fields" and "decision makers".

For the moment, the Design-time step does not allow automatic design of the event subscription, sending and reception in our collaborative processes, as they are not taken into account in the process cartography. We have added all the event subscriptions, sending and reception by hand at the workflow level. We obviously plan to integrate the event definition and use into the Design-time steps in the future.

Another limitation of our work is due to the nature of the system components: we are based on organizations' Information Systems, and are thus technically dependent on the network. If the network goes down, a part or, in the worst case, the entire MIS fails too. Hardware security measures should be taken to protect the physical network, in addition to the security measures taken to protect the data network against acts of piracy or unreliable data. Another security point to deal with is the access to the events: which sources of events, or types of events can be considered trustworthy? Which ones are allowed to provide events in our network? We need to envisage a governance tool to manage subscriptions to event types.

Based on the considerations in the previous sections, we can finally conclude by describing the two main features of an agility service, based on events and complex event processing, integrated into the MIS infrastructure:

- *Detection*: considering the fact that events may be used to update models of the collaborative situation (events are viewed as sources of context), the adaptation service would be in charge of maintaining both a field model and an expected model. Both these models would describe not only the considered crisis situation but also the actors involved. The *field* model would be based on events coming from sensors, measurements and reports (generally any concrete information) and the *expected* model would be based on events coming mainly from workflow orchestration. Measuring the distance between both these *field* and *expected* models, an adaptation service could be able to detect any critical divergence between the supposed situation and the real one, hence justifying a need for adaptation.

- *Adaptation*: once such a need for adjustment has been detected, the adaptation service could use the updated *field* model in order to restart any of the MISE steps (collaborative process deduction, semantic reconciliation or deployment), depending on the nature of the evolution. The nature of the divergence could be diagnosed according to the distance between *field* and *expected* models (insofar as it is a "multi-dimensions distance") and could hence help to define which step should be restarted.

Both the main components of agility (*detection* and *adaptation*) might thus be covered by this agility service. Besides, the model-driven nature of the MISE system would also provide *responsiveness* in order to ensure complete agility (because most of the transitions are automated) of the MIS.

Furthermore, this adaptation service could also provide a very interesting complementary feature, which is currently a promising perspective:

- *Watching*: based on the previously described features, one can easily imagine that, considering for instance a predefined area (in a crisis management context, this would be a geographical area, but it might also be a business area in another context) such a service could gather events coming from that area (from sensors or any connected component / person). These events could be used to create and update a living image of the observed area (*i.e.* a model of the considered system). Relevant evolutions of that model could be exploited for diagnostic purposes if there is a sudden change and potentially a need to use this current model of the situation to deduce a collaborative response. Finally, this watching function could allow crisis detection and the beginning of a crisis response based on up-to-date and automatically-gathered knowledge.

## 5. Acknowledgements

The MISE 2.0 research work has been partially funded by the French Research Agency (ANR) regarding the research project SocEDA *(SOCial Event Driven Architecture)* (Grant ANR-10-SEGI-013) and the European Commission under Seventh Framework Program (FP7) regarding the research project PLAY *(Pushing dynamic and ubiquitous interaction between services Leveraged in the Future Internet by ApplYing complex event processing)* (Grant FP7-258659).

SocEDA aims to provide dynamic and adaptive workflows to collaborative situations through EDA and CEP. PLAY aims to develop and validate an elastic and reliable architecture for dynamic and complex, event-driven interaction in large highly distributed and heterogeneous service systems.

The authors would like to thank the project partners for their advice and comments regarding this work.

*collaborative Business Process Deduction is a methodology of Mediator Modeling*